Journal of Economics and Financial Analysis, Vol:2, No:1 (2018) 129-149

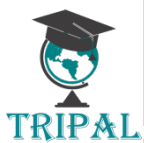

**Journal of Economics and Financial Analysis**

Type: Double Blind Peer Reviewed Scientific Journal
Printed ISSN: 2521-6627 | Online ISSN: 2521-6619
Publisher: Tripal Publishing House | DOI:10.1991/jefa.v2i1.a15

Journal homepage: www.ojs.tripaledu.com/jefa

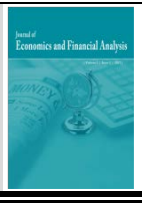

# Does it take two to tango: Interaction between Credit Default Swaps and National Stock Indices

Yhlas SOVBETOV[*,a], Hami SAKA[b]

[a] *Department of Economics, London School of Commerce, United Kingdom*
[b] *Department of Economics, Istanbul University, Turkey*

**Abstract**

*This paper investigates both short and long-run interaction between BIST-100 index and CDS prices over January 2008 to May 2015 using ARDL technique. The paper documents several findings. First, ARDL analysis shows that 1 TL increase in CDS shrinks BIST-100 index by 22.5 TL in short-run and 85.5 TL in long-run. Second, 1000 TL increase in BIST index price causes 25 TL and 44 TL reducation in Turkey's CDS prices in short- and long-run respectively. Third, a percentage increase in interest rate shrinks BIST index by 359 TL and a percentage increase in inflation rate scales CDS prices up to 13.34 TL both in long-run. In case of short-run, these impacts are limited with 231 TL and 5.73 TL respectively. Fourth, a kurush increase in TL/USD exchange rate leads 24.5 TL (short-run) and 78 TL (long-run) reductions in BIST, while it augments CDS prices by 2.5 TL (short-run) and 3 TL (long-run) respectively. Fifth, each negative political events decreases BIST by 237 TL in short-run and 538 TL in long-run, while it increases CDS prices by 33 TL in short-run and 89 TL in long-run. These findings imply the highly dollar indebted capital structure of Turkish firms, and overly sensitivity of financial markets to the uncertainties in political sphere. Finally, the paper provides evidence for that BIST and CDS with control variables drift too far apart, and converge to a long-run equilibrium at a moderate monthly speed.*

**Keywords:** *Credit Default Swaps (CDS); BIST-100 index; Cointegration; ARDL.*

**JEL Classification:** *E00, E44.*

---

[*] Corresponding author. Office CH-28, London School of Commerce, Chaucer House, London, SE1 1NX, London, UK. Tel: +44 207 357 0077 / ext.364
E-mail addresses: ihlas.sovbetov@lsclondon.co.uk (Y. Sovbetov), hamisaka@gmail.com (H. Saka)

## 1. Introduction

Doubtlessly the last global financial crisis has put a spotlight on Credit Default Swaps (CDS), a contract between parties where buyer transfers the risk of default to the seller of swap, which has been utilized as primary hedging technique for both corporate and sovereign credit risks since early 1990s. Since its introduction in 1994 by JP Morgan, it rapidly has integrated into the daily life of many traders, regulators, and financial economists. These swaps can be used for several purposes. For instance, risk managers might use them to manage and hedge specific credit risks, while investors might utilize them to earn higher returns by buying out additional risks. Traders might also use them to make profit from bid-offer spreads, while some other agents might wish to get the advantage of tax arbitrage with these derivatives.

The CDS market has been experiencing rapid grows since mid 1990s (the date when the first transactions were facilitated) reaching the market size of 900 billion USD in the eve of 2000. Further by the end of 2005 it scaled up to 14 trillion USD, and continued its rapid growth (hitting about 60 trillion USD level) until recent sub-prime crisis in 2007-2008. Henceforth the market size experienced continuous decreases. Just in the first year of the crisis, it shrank almost by 30% melting down to the level of 41 trillion USD, and the following years it continued to decrease but at a slower rate (see figure 1). Today, the size of credit derivates market is about 14.5 trillion USD, and we all have learned inevitable role of these swaps by experiencing its (direct and indirect) macro- and micro-economic impacts. Yet, although thousands of studies discuss these economic roles of CDS, some key issues are still hotly debated and remain controversial.

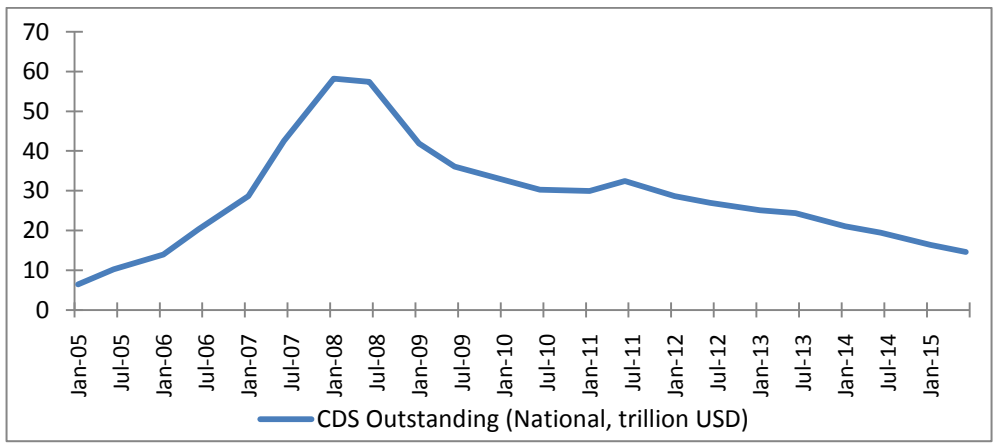

**Figure 3.** CDS Outstanding (National, trillion USD)

With this motivation, we examine how CDS prices and BIST-100 stock index interact in short- and long-run over January 2008 to May 2015 using the largest available data set of Turkey's sovereign CDS spreads. Our findings show that the impact running from monthly changes in BIST-100 index to fluctuations in CDS is higher than the vice versa impact. Plus, we also document that both BIST and CDS are severe responsive to changes in TL/USD exchange rate and political instability both in short- and long-run. These findings address to the highly dollar indebted capital structure of Turkish firms, and overly sensitivity of financial markets to the uncertainties in political sphere. We also observe limited impact running from interest and inflation rate on to BIST and CDS separately.

The remainder of this paper is structured as following. Next section briefly reviews related literature. Third section covers data description and methodology. Fourth section provides key findings with their interpretations, and the final section concludes.

**2. Literature Review**

A sizeable literature exists on determinants of CDS spreads such as Longstaff et al. (2003), Galil et al. (2014), and Norden and Weber (2004). These studies explore the association between CDS spreads and stock returns, bond yields, credit ratings, interest rates and other financial indicators. Galil et al. (2014) present three variables to explain changes in CDS spreads: stock return, the change in stock return volatility and the change in the median CDS spread in the rating class. They use dataset of 718 US firms during the period from 2002 to 2013 and obtain parallel results with Blanco et al. (2005) and Trutwein et al. (2011). They also observe that the last Global Financial Crisis caused a structural change in CDS spreads.

More specifically, Longstaff et al. (2003) consider a VAR model for explaining the lead-lag relationships between stock returns, CDS, and bond spreads with weekly data in US, and report that stock returns and CDS spreads lead to the bond spreads. Norden and Weber (2009) also state that stock price indices in prior can predict the CDS and bond spreads. Similarly, Zhu (2006) examines how CDS spreads associated with bond markets, and finds that CDS spreads provide more accurate measurement of default risk than bond spreads. In addition to this, there are some studies which argue that there is a negative relationship among interest rate variables and CDS spreads such as Fama (1984) and Estrella and Hardouvelis (1991).

The relationships between Credit Default Swaps and Credit Ratings are analyzed in various studies. CDSs create more sensitive impression than Credit

Ratings do. It can react to the development on the market more quickly and stimulate the markets in advance as supported by the results that were concluded from previous studies. And of course, here, the operation of the mechanism of Credit Rating measurement and CDS are crucial. Whilst CDSs are announced continuously and daily, rating changes are made in the longer term. Besides, rating scale is not sensitive because it does not include wide range as CDS. Coronado et al. (2011) analyze relationship between stock returns and sovereign CDS spreads by sampling 8 European countries between the periods of 1$^{st}$ January 2007 to 30$^{th}$ July 2010. They show that CDS spread changes are negative correlated with stock returns over the sample countries. However, the significance of correlations is more severe in Italy, Greece, Spain, and Portugal. They point out that these countries have higher risk premiums. Furthermore, they conduct a VAR model to explain the lead-lag link among sovereign CDS spreads and stock returns, and find that changes in stock returns lead sovereign CDS spreads changes between January 2007 to December 2009 periods, and vice versa for during the period of January 2010 to July 2010. They partly disagree with Fung et al. (2008) and Narayan et al. (2014) who bind bidirectional interaction of stock prices and CDS to few conditions. Briefly, Fung et al. (2008) report that stock price affects CDS spreads of firms that have investment grade rating, and vice versa impact is valid only for firms with high payment capacity. On the other hand, Narayan et al. (2014) examine bidirectional interaction of stock prices and CDS concentrating on firms' industries, and find that the interaction exists only in energy, finance, materials, consumer discretion, health care, and industrials sectors.

On the other hand, extant literature (Norden and Weber, 2004; Hull et al., 2004; Finnerty et al., 2013) shows that Credit Ratings are another chief determinant of CDS prices. Hull et al. (2004) assert that review for downgrade ratings forecasts majority of changes in CDS in advance. Norden and Weber (2004) find this time lag (between rating downgrades and changes in CDS spreads) as 60-90 days. Similarly, Ismailescu and Kazemi (2010) report that CDS spreads are significantly sensitive to the rating downgrades. They observe that negative information about credit ratings, i.e. rating downgrades, predicts 42.6% of CDS spreads, however, positive information (rating upgrades) has a limited impact.

**3. Data and Methodology**

To our knowledge, the literature about interaction between Turkey's CDS and stocks is limited. To resolve this insufficiency we investigate their bi-directional over the period of 2008-2015 both with daily and monthly data. In our daily analysis, following Coronado et al. (2011) we establish VAR model as below.

$$VAR(p) = \Delta BIST_t = \alpha_0 + \sum_{i=1}^{p} \alpha_i \Delta BIST_{t-i} + \sum_{i=1}^{p} \beta_i \Delta CDS_{t-i} + e_{1t} \qquad (1a)$$

$$VAR(p) = \Delta CDS_t = \beta_0 + \sum_{i=1}^{p} \beta_i \Delta CDS_{t-i} + \sum_{i=1}^{p} \alpha_i \Delta BIST_{t-i} + e_{2t} \qquad (1b)$$

where *p* is optimal lag length; *ΔBIST$_t$* is a return of BIST-100 index, and *ΔCDS$_t$* is a change credit default swap rate at day *t*.

On the other hand, in monthly analysis, we consider those two variables (BIST and CDS) in their raw money units, and augment the model by adding control variables such as inflation (*CPI*), exchange (*EX*), and interest rates (*INT*). Besides we also introduce a dummy (*DPOL*), which equates to 1 for the month when significant political events, that distress the financial markets, are occurred, and turns back to zero for other months, to capture the affects of political instability as below.

$$BIST_t = \alpha_0 + \sum_{i=1}^{p} \alpha_i BIST_{t-i} + \sum_{i=1}^{p} \beta_i CDS_{t-i} + \sum_{i=1}^{4,p} \theta_{it} \psi_{it} + \omega_{1t} \qquad (2a)$$

$$CDS_t = \beta_0 + \sum_{i=1}^{p} \beta_i CDS_{t-i} + \sum_{i=1}^{p} \alpha_i BIST_{t-i} + \sum_{i=1}^{4,p} \theta_{it} \psi_{it} + \omega_{2t} \qquad (2b)$$

where *BIST$_t$* is a price of BIST-100 index in units of Turkish Lira (TL), and *CDS$_t$* is a price of credit default swaps in units of TL at month *t*. The "$\psi$" stands for aforementioned controls, where inflation and interest rates are proxied by consumer price index levels (*CPI*) and short-term lending rates of Turkish Central Bank respectively. Besides the dummy (*DPOL*) accounts the most significant political crisis alongside with five electoral events during January 2008 and May 2015 which are referred in Appendix-A.

More specifically, the table 1 displays descriptive statistics of input variables where data for *BIST*, *CDS*, and *CPI* are gathered from Eikon datastream of Thomson Reuters, meanwhile interest and exchange rates are obtained from OECD and OANDA databases respectively.

**Table 1.** Descriptive Analysis

|                   | BIST  | CDS    | EX   | CPI   | INT   | DPOL   |
|-------------------|-------|--------|------|-------|-------|--------|
| *Mean*            | 60.27 | 225.31 | 1.74 | 8.12  | 7.24  | 0.06   |
| *Median*          | 62.36 | 195.76 | 1.76 | 8.17  | 6.50  | 0.00   |
| *Maximum*         | 88.95 | 487.65 | 2.65 | 12.06 | 16.75 | 1.00   |
| *Minimum*         | 24.03 | 119.66 | 1.17 | 3.99  | 1.50  | 0.00   |
| *Std. Dev.*       | 17.09 | 85.58  | 0.33 | 1.85  | 4.22  | 0.23   |
| *Skewness*        | -0.36 | 1.58   | 0.53 | -0.11 | 1.01  | 3.83   |
| *Kurtosis*        | 2.30  | 4.93   | 2.96 | 2.53  | 3.20  | 15.66  |
| *Jarque-Bera stat.* | 3.69 | 50.42 | 4.05 | 1.00  | 15.04 | 802.72 |
| *Probability*     | 0.16  | 0.00   | 0.13 | 0.61  | 0.00  | 0.00   |
| *Observations*    | 88    | 88     | 88   | 88    | 88    | 88     |

**Notes:** *BIST* is a price of BIST-100 index in thousands of Turkish Lira (TL); *CDS* is a price of credit default swaps in TL; *EX* is a TL/USD exchange rate; *CPI* is a consumer price index level; *INT* is a short-term lending rate announced by Turkish Central Bank; *DPOL* stands for significant political events that distress the Turkish financial market.

### 3.1. Model Specification

Initially we analyze the characteristics of all series under Augmented Dickey-Fuller (ADF) test, formulated as below, lest they violate the stationarity assumption of OLS.

$$\Delta \Omega_t = \theta_0 + \theta_1 T + \rho \Omega_{t-1} + \sum_{i=2}^{k} \theta_i \Delta \Omega_{t-1} + \varepsilon_t$$

where $\Delta\Omega_t$ is the first difference of a variable $\Omega$; *T* is a trend, and $\theta_1$ is its multiplier; *k* is a optimal lag length; and $\varepsilon_t$ is White Noise residual term. Here, ADF hypothesizes $H_0$ $(\rho=0)$ against alternative $(\rho\neq 0)$. Failure of rejection of null hypothesis indicates that the variable satisfies the stationarity assumption of OLS.

The table 2 presents results of ADF test where all series except *DPOL* appear non-stationary at level. But they can be converted to a stationary through differencing methodology. As a result, we conclude that only *DPOL* is an *I(0)* variable, while others are *I(1)*. As a result the input series are not integrated at same degree, therefore we adjust our (2) VAR model into an Autoregressive Distributed Lag (ARDL) model as in (2.1) that tests long-run relationship (cointegration) of *I(0)* and *I(1)* series.

$$\Delta BIST_t = \delta_0 + \sum_{i=1}^{p} \delta_i \Delta BIST_{t-i} + \sum_{i=0}^{q} \theta_i \Delta CDS_{t-i} + \sum_{j=1}^{4} \sum_{i=0}^{k,l,m,n} \eta_{ji} \Delta \psi_{jt-i}$$

$$+ \varphi_1 BIST_{t-1} + \varphi_2 CDS_{t-1} + \sum_{j=1}^{4} \varphi_{3j} \psi_{jt-1} + \mu_{1t} \quad (2.1a)$$

$$\Delta CDS_t = \theta_0 + \sum_{i=1}^{p'} \theta_i \Delta CDS_{t-i} + \sum_{i=0}^{q'} \delta_i \Delta BIST_{t-i} + \sum_{j=1}^{4} \sum_{i=0}^{k',l',m',n'} \eta_{ji} \Delta \psi_{jt-i}$$

$$+ \varphi_1 CDS_{t-1} + \varphi_2 BIST_{t-1} + \sum_{j=1}^{4} \varphi_{3j} \psi_{jt-1} + \mu_{2t} \quad (2.1b)$$

Here, *p, q, k, l, m, n* and their primes are optimal lag lengths for related variables in the model that are determined by Schwarz Information Criterion (SIC), and $\mu_t$ and $\mu_t$ are White noise stationary residual terms. Indeed, this methodology is also known as bound testing approach that is pioneered by Pesaran et al. (2001), where the null hypothesis of $\varphi_i=0$ is tested against $\varphi_i \neq 0$ with Wald test considering critical lower and upper bound values. The case, where Wald F-statistics is below lower bound, indicates that series are not cointegrated and there is no any long-run relationship between them. In fact this association exists only if the Wald F-statistics exceeds the critical upper bound, and in case it falls between bounds then the cointegration is inconclusive.

**Table 2.** Output of ADF Analysis

| Variables | Level | | | 1st Difference | | |
|---|---|---|---|---|---|---|
| | Prob. | Lag | DW | Prob. | Lag | DW |
| **BIST** | 0.7502 | 0 | 1.9201 | 0.0000 | 0 | 1.9554 |
| **CDS** | 0.2139 | 0 | 1.7684 | 0.0000 | 0 | 2.0209 |
| **EX** | 0.9941 | 1 | 1.9308 | 0.0000 | 0 | 1.9316 |
| **CPI** | 0.0531 | 0 | 1.5998 | 0.0000 | 0 | 2.0088 |
| **INT** | 0.2716 | 1 | 2.0723 | 0.0000 | 0 | 2.0662 |
| **DPOL** | 0.0000 | 0 | 2.0079 | 0.0000 | 1 | 1.9687 |

**Notes:** *DW* is Durbin-Watson statistics. The lag is automatically determined by Schwarz Information Criterion (SIC) with maximum 8 lags.

Detection of long-run cointegration in (2.1) equation emerge possibililty of following short-run interaction alongside with vector error-correction model (VECM) that can be formulated as below.

$$\Delta BIST_t = \delta_0 + \sum_{i=1}^{p} \delta_i \Delta BIST_{t-i} + \sum_{i=0}^{q} \theta_i \Delta CDS_{t-i} + \sum_{j=1}^{4} \sum_{i=0}^{k,l,m,n} \eta_{ji} \Delta \psi_{jt-i}$$
$$+ \lambda_1 ECT_{1,t-1} + v_{1t} \qquad (2.2a)$$

$$\Delta CDS_t = \theta_0 + \sum_{i=1}^{p'} \theta_i \Delta CDS_{t-i} + \sum_{i=0}^{q'} \delta_i \Delta BIST_{t-i} + \sum_{j=1}^{4} \sum_{i=0}^{k',l',m',n'} \eta_{ji} \Delta \psi_{jt-i}$$
$$+ \lambda_2 ECT_{2,t-1} + v_{2t} \qquad (2.2b)$$

where $ECT_1$ ($\omega_1$) and $ECT_2$ ($\omega_2$) are stationary residual of (2a) and (2b) respectively, and $\lambda_1$ and $\lambda_2$ are their multiplier that are expected to be significant and between -1 and 0 for robustness of VECM model.

## 4. Analysis and Findings

### 4.1. Examination with daily data

Recalling (1), we examine daily interaction between return of BIST-100 index and change in *CDS*, both series are stationary at level, by employing Granger (1969) causality test. Here the optimal lag length *"p"* is determined with information criterion tests as shown in table 3.

**Table 3.** Lag Length Selection Test

| Lag | LR | FPE | AIC | SC | HQ |
| --- | --- | --- | --- | --- | --- |
| 0 | NA | 2.26e-07 | -9.628784 | -9.622192 | -9.626339 |
| 1 | 48.21678 | 2.20e-07 | -9.653375 | **-9.633600*** | **-9.646041*** |
| 2 | 5.636000 | 2.20e-07 | -9.651943 | -9.618985 | -9.639720 |
| 3 | 12.73105 | **2.20e-07*** | **-9.654863*** | -9.608722 | -9.637750 |
| 4 | 5.898214 | 2.20e-07 | -9.653601 | -9.594276 | -9.631598 |
| 5 | 1.080703 | 2.21e-07 | -9.649383 | -9.576876 | -9.622491 |
| 6 | **15.49522*** | 2.20e-07 | -9.654032 | -9.568341 | -9.622251 |

**Notes:** The lag length criteria test shows comparative outcome of selection criteria. LR is sequential modified LR test statistic (each test at 5% level); FPE is Final prediction error; AIC is Akaike information criterion; SC is Schwarz information criterion; and HQ is Hannan-Quinn information criterion. The asterisks (*) indicates lag order selected by the criterion.

The results show that FPE and AIC point the lag 3 as optimal lengths while SIC and HQ suggest the lag 1. In this case, we decide to employ the Granger (1969) causality test both with lag 1 and 3 separately as in table 4 where the both cases find that there is bi-directional causality between *ΔBIST* and *CDS*.

**Table 4.** Lag Length Selection Test

| Model | Null Hypothesis | F-Stat. | Prob. | Observation |
|---|---|---|---|---|
| VAR(1) | $\Delta CDS$ does not Granger cause $\Delta BIST$ | 7.3300 | 0.0069 | 1644 |
| | $\Delta BIST$ does not Granger cause $\Delta CDS$ | 3.1344 | 0.0194 | |
| VAR(3) | $\Delta CDS$ does not Granger cause $\Delta BIST$ | 2.7865 | 0.0395 | 1642 |
| | $\Delta BIST$ does not Granger cause $\Delta CDS$ | 6.7360 | 0.0002 | |

The Granger causality results reveal that *BIST* and *CDS* are tightly associated with each other. More specifically, the result of VAR(1) model, in table 5, refers that a unit increase in today's *ΔCDS*, indeed, leads tomorrow's *ΔBIST* decrease by -0.04 units. The VAR(1) model where $\Delta BIST_t$ is dependent variable suffers from heteroscedasticity problem in residuals. By assigning White heteroscedasticity consistent coefficient covariance we fix it. As a result we get slightly different standard errors (which we can trust), and asymptotically standard normal distributed t-statistics where significance of $\Delta CDS_{t-1}$ has decreased from 1% to 5% level.

On the other hand, VAR(1) model, where *ΔCDS* is dependent, appears perfectly healthy. It estimates -0.1669 units impact from *ΔBIST* to *ΔCDS* which is actually triple of *ΔCDS-ΔBIST* (-0.0369). Hereby, we conclude that a unit increase in today's *ΔBIST* shrinks *ΔCDS* by 0.17% in following day.

The right-hand side of the table 5 displays results of VAR(3) where both *ΔBIST* and *ΔCDS* dependent models suffer from serial correlated and heteroscedastic residuals. We could cure them, only, by removing third lags of dependent variables from the models. Meantime, we assign White heteroscedasticity consistent coefficient covariance, and get new robust models (*ΔBIST"* and *ΔCDS"*) that estimate quite similar coefficient as VAR(1), -0.03 units impact from *ΔCDS* to *ΔBIST* and -0.17 units impact from *ΔBIST* to *ΔCDS*.

**Table 5.** Results of VAR Analysis

| Variable | VAR(1) | | | VAR(3) | | | |
|---|---|---|---|---|---|---|---|
| | ΔBIST$_t$ | ΔBIST$_t$' | ΔCDS$_t$ | ΔBIST$_t$ | ΔBIST$_t$" | ΔCDS$_t$ | ΔCDS$_t$" |
| ΔBIST$_{t-1}$ | 0.1252*** (0.0155) | -0.1252** (0.0128) | -0.0769*** (0.0074) | 0.1216*** (0.0171) | 0.1266*** (0.0185) | -0.0426*** (0.0057) | -0.0408*** (0.0051) |
| ΔBIST$_{t-2}$ | - | - | - | 0.0758* (0.0426) | 0.0719** (0.0364) | -0.0138* (0.0075) | -0.0142* (0.0081) |
| ΔBIST$_{t-3}$ | - | - | - | 0.0163** (0.0080) | - | -0.0106*** (0.0026) | -0.0173** (0.0082) |
| ΔCDS$_{t-1}$ | -0.2369*** (0.0499) | -0.2369** (0.0584) | 0.3776*** (0.0307) | -0.2158*** (0.0392) | -0.2230*** (0.0473) | 0.3532*** (0.0313) | 0.3654*** (0.0385) |
| ΔCDS$_{t-2}$ | - | - | - | -0.1154** (0.0562) | -0.1042** (0.0495) | 0.1221*** (0.0278) | 0.1309*** (0.0255) |
| ΔCDS$_{t-3}$ | - | - | - | -0.0449* (0.0260) | -0.0135 (0.0088) | 0.0386** (0.0192) | - |
| C | 0.0008* (0.0005) | 0.0008* (0.0005) | 0.0004 (0.0003) | 0.0008** (0.0004) | 0.0008** (0.0004) | 0.0005 (0.0004) | 0.0004 (0.0004) |
| R$^2$ | 0.4062 | 0.3658 | 0.3875 | 0.5931 | 0.5358 | 0.6474 | 0.6161 |
| DW | 1.9897 | 1.9897 | 1.9853 | 1.1979 | 2.0170 | 1.1390 | 1.9977 |
| BG LM | 0.6131 | 0.6131 | 0.9873 | 0.0001 | 0.6973 | 0.0001 | 0.4695 |
| BPG Test | 0.0197 | - | 0.8517 | 0.0000 | - | 0.0000 | - |

**Notes:** The numbers in upper part of the table are coefficients estimated by VAR analisys technique where significance levels follows as *:10%, **:5%, and ***:1%. The bottom part of the table shows diagnostics of the VAR models. BG LM is Breusch-Godfrey Serial Correlation LM test with H$_0$: residuals of the model are not serially correlated. BPG is Breusch-Pagan-Godfrey heteroscedasticity test with H$_0$: residuals of the model are homoskedastic. Both tests reports chi-square probabilities. DW is Durbin-Watson statistics.

### 4.2. Monthly Analysis

Further, we investigate both short-run and long-run interaction of *BIST* and *CDS* holding *EX, CPI, INT,* and *DPOL* as control variables. Initially, we present raw regression analysis results in the table 6, where all variables are in first-differenced form (as they are not stationary at level) except *DPOL*. Besides, table 6 shows cross series correlation of all input variables. Although correlation coefficients seem to be low to question presentece of collinearity problem, one might think that the correlation of -0.5339 among inflation and interest rate imply collinearity. Lest this potential collinearity motivated by correlation between inflation rate (*CPI*) and interest rate (*INT*), we do not include these two variables in same model simultaneously. Instead, we test their contribution separately.

**Table 6.** Correlatin among Series

|      | BIST    | CDS     | EX      | CPI     | INT     | DPOL   |
|------|---------|---------|---------|---------|---------|--------|
| *BIST* | 1       | -0.3952 | 0.6961  | -0.1928 | -0.6519 | 0.0192 |
| *CDS*  | -0.3952 | 1       | 0.2781  | 0.4001  | 0.4717  | 0.2643 |
| *EX*   | 0.6961  | 0.2781  | 1       | -0.0504 | -0.3745 | 0.1702 |
| *CPI*  | -0.1928 | 0.4001  | -0.0504 | 1       | 0.5339  | 0.0979 |
| *INT*  | -0.6519 | 0.4717  | -0.3745 | 0.5339  | 1       | 0.0592 |
| *DPOL* | 0.0192  | 0.2643  | 0.1702  | 0.0979  | 0.0592  | 1      |

Table 7 reports OLS estimations of *ΔBIST*-dependent and *ΔCDS*-dependent models in each column. The *ΔBIST*-dependent model estimates a significant negative short-run impact from *ΔCDS* to *ΔBIST* (-0.02) at 1% significance level. It indicates that 1 TL increase in *ΔCDS* price causes BIST drop by 17.3 TL. Exchange rate and interest rate seem to have negatively related with BIST as well. A *"kurush"* (1% of TL) increase in exchange rate shrinks BIST by 56 TL at 1% significance level, while 1 bps increase interest rate leads 312 TL reduction in BIST at 10% significance level. More importantly, dummy variable *DPOL* also derives statistically significant negative estimate of -0.25 at 1% level. However, this model violates non-serially correlated and homoscedastic residuals assumption of OLS. By dropping intercept term and assigning White heteroscedasticity consistent coefficient covariance we fix this problem for the sake of slight decrease in significance of *ΔCDS* (from 1% to 5%) and slight changes in other variables magnitudes, i.e. *ΔCDS* and *ΔINT* increase (in absolute value) to -0.0220 and -0.3388 respectively, while *ΔEX* and *DPOL* shrinks to -5.04 and -0.2416 respectively. When we replace interest rates with inflation rates, the model derive fairly similar estimates, however, newly included inflation fail to be statistically significant. This clearly shows *ΔINT-ΔCPI* trade-off is in favor of *ΔINT* as role of inflation over *ΔBIST* is insignificant.

On the other hand, *ΔCDS* dependent model appears perfectly healthy. By accounting 58.53% variations in *ΔCDS*, the model predicts significant negative impact running from *ΔBIST* to *ΔCDS* (-35.73 in *ΔINT* case and -36.93 in *ΔCPI* case) at 1% significance level. This indicates that 1000 TL increase in BIST index reduces *ΔCDS* price by 37 TL. Besides, a strong positive impact running from exchange rate to CDS is also documented at 1% significance level, indicating a kurush (1/100 TL) increase in exchange rate scales *ΔCDS* up by 2.93 TL (in *ΔINT* case) or 3.09 TL (in *ΔCPI* case). Interestingly, inflation becomes a significant explanatory variable of *ΔCDS* that contribute by additional 5% of $R^2$ value, while interest rate fails to be statistically significant. 1 bps increase in *Δ*CPI causes *Δ*CDS increase by 4.57 TL.

Plus, the coefficient of *DPOL* is predicted as -36 at 1% significance level. It indicates that each political uncertainty scales the CDS prices up by 36 TL.

**Table 7.** Results of OLS Analysis

| Variable | (1) ΔBIST | (2) ΔBIST' | (3) ΔBIST | (4) ΔCDS | (5) ΔCDS |
|---|---|---|---|---|---|
| ΔCDS | -0.0173*** (0.0044) | -0.0220** (0.0111) | -0.0178*** (0.0051) | - | - |
| ΔBIST | - | - | - | -35.7381*** (8.5139) | -36.9325*** (8.5495) |
| ΔEX | -5.6057*** (1.2033) | -5.0388*** (1.1206) | -5.1926*** (1.2122) | 292.84*** (81.0982) | 308.93*** (83.7342) |
| ΔCPI | - | - | -0.0831 (0.4365) | - | 4.5738** (2.2560) |
| ΔINT | -0.3120* (0.1661) | -0.3388* (0.1835) | - | 6.7281 (6.6600) | - |
| DPOL | -0.2472*** (0.0762) | -0.2416*** (0.0805) | -0.2423*** (0.2969) | 35.6275*** (6.0651) | 35.8889*** (6.1202) |
| C | 5.3268*** (1.4732) | - | 1.3676*** (0.3613) | -9.0147*** (2.3064) | -9.9546*** (2.5137) |
| $R^2$ | 0.5512 | 0.5127 | 0.5485 | 0.5511 | 0.5853 |
| DW | 2.5937 | 2.0267 | 2.0584 | 2.0912 | 2.0675 |
| BG LM | 0.0114 | 0.9238 | 0.2155 | 0.3524 | 0.3128 |
| BPG | 0.0024 | - | 0.1372 | 0.1569 | 0.1974 |
| White | 0.0000 | - | 0.1190 | 0.2680 | 0.2680 |

**Notes:** The numbers in upper part of the table are coefficients estimated by OLS analisys technique where significance levels follows as *:10%, **:5%, and ***:1%. Note that BIST is in thousands TL unit (000 TL); CDS is in TL; CPI and INT are in percentage; EX is in raw currency rate. The bottom part of the table shows diagnostics of the OLS models. BG LM is Breusch-Godfrey Serial Correlation LM test with $H_0$: residuals of the model are not serially correlated. BPG is Breusch-Pagan-Godfrey heteroscedasticity test with $H_0$: residuals of the model are homoskedastic. Both tests reports chi-square probabilities. DW is Durbin-Watson statistics.

Subsequently we estimate the long-run tango of BIST and CDS by recalling (2). Initially, we determine *p, q, k, l, m, n* and their primes by employing optimal lag length selection tests in Eviews 9.0 software. The figure 2 displays the output of this analysis where, following minimum AIC value, lag lengths are specified as *p=2, q=1, k=1, l=0, m=0,* and *n=0* for ARDL model (2.1a). In case of (2.1b), both AIC and SIC finds the lag lengths as *p'=1, q'=1, k'=1, l'=1, m'=0,* and *n'=0*.

Further, we check diagnostics of our selected ARDL models for BIST and CDS separately. We use HAC-robust standard error to fix potential serial correlation and heteroscedasticity problems. Plus, we check check stability of these models with Cumulative Sum of the Recursive Residuals (CUSUM) test to make sure that they do not involve any structural breaks. We show results of diagnostic analysis in figure 3 where we document that both models are stable over time as their CUSUM (blue) lines remains between ±5% significance (two red) lines.

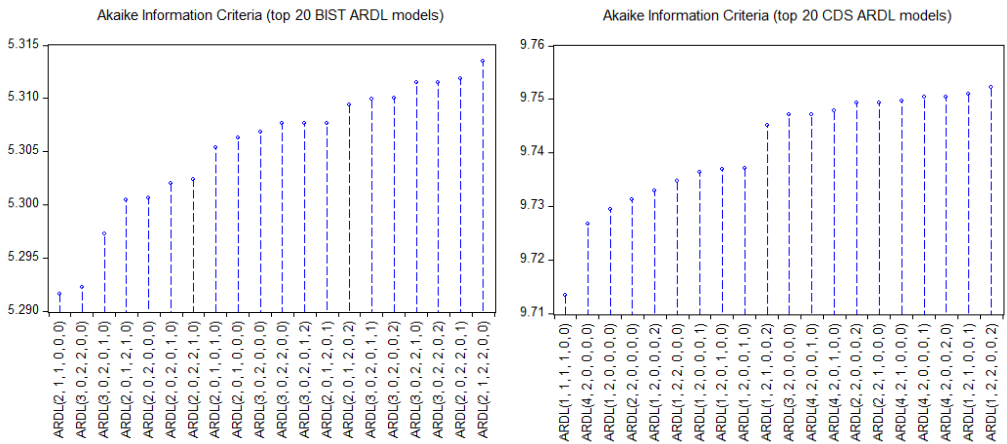

**Figure 2.** ARDL Lag Specification for BIST and CDS models.

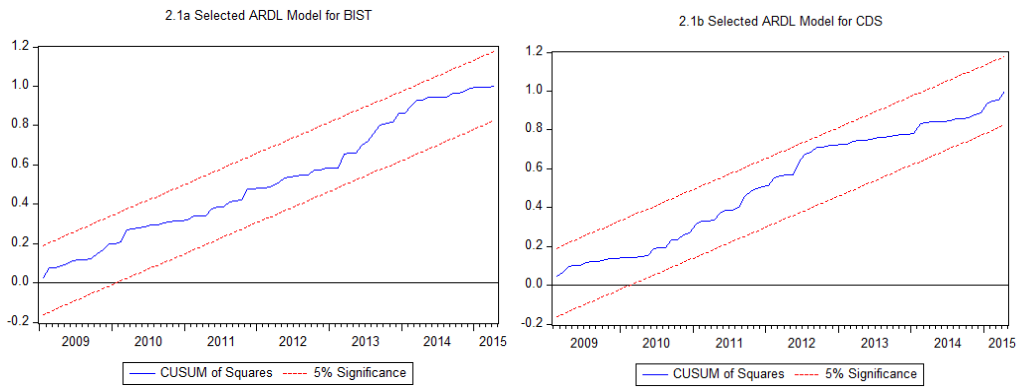

**Figure 3.** Stability of Selected BIST and CDS ARDL Models

Further, we derive estimates for these ARDL models in table 8 (Panel A) where first column presents estimates for equation 2.1a accounting 45.49% variations in *ΔBIST*. We hypothesize $\varphi_i=0$ statement for both models utilizing Wald test. Its result is reported at the bottom part of the table 8 where model 2.1a derives t-value of 4.39 which highly exceeds Pesaran et al. critical value 1% significance level, indicating existence of long-run cointegration between series. We dear critical values of Case I that are presented in Pesaran et al (2001) at table CI. Because, case I is specified for models that do not comprise intercept value and any kind of trends, and so both of our models do not include intercept and trend factors as they appear statistically insignificant.

The negative ratio of coefficients to dependent variable ($-\varphi_i/\varphi_6$) shall show the direction and magnitude of their long-run relationship. Panel B of table 8 shows long-run multiplier derived by Panel A estimation. On this basis, the *ΔBIST*-dependent model implies that 1 TL increase in CDS prices leads BIST-100 index to decrease by 85 TL at 1% significance level in long-run. Equally, the model also forecasts that a kurush (1% of TL) value lose of TL against USD, leads an average of 78 TL decrease in BIST-100 index value in long-run. It might be addressed to the highly dollar indebtness structure of Turkish firms, so that the risk of going default increases with the increase of debt that is triggered by appreciation of dollar against TL. More interestingly, 1 base point increase in interest rate causes BIST-100 index decrease by 359 TL in long-run, while each stressful political event shrinks BIST-100 index by roughly 538 TL in long-run, which makes sense, as uncertainties in regulative power negatively affect the financial markets.

The second column of panel A displays estimates of equation 2.1b where *ΔCDS* is dependent variable. The model's long-run multipliers in panel B indicate that 1000 TL increase in BIST index value causes 44 TL reduction in CDS prices in long-run. Likewise, a kurush increase in exchange rate (TL/USD) augments CDS prices nearly by 3 TL at 1% significance level in long-run. One percentage point increase in *CPI* will enhance CDS prices by 13.34 TL, while each political tensions add 89 TL into CDS prices in long-run at 1% significance level.

**Table 8.** Results of Bound Test and Long-run Multipliers

| PANEL A | (2.1a) ΔBIST | (2.1b) ΔCDS | PANEL B | Long-run ΔBIST | Long-run ΔCDS |
|---|---|---|---|---|---|
| $\Delta BIST_t$ | - | -6.0800*** (1.5238) | - | - | - |
| $\Delta BIST_{t-1}$ | -0.2951*** (0.0941) | -3.5304** (1.6235) | - | - | - |
| $\Delta EX_t$ | -5.9528*** (4.2677) | 309.4381** (145.7939) | - | - | - |
| $\Delta CPI_t$ | 0.1788 (0.4334) | - | - | - | - |
| $\Delta CPI_{t-1}$ | -0.0879** (0.0438) | - | - | - | - |
| $CDS_{t-1}$ ($\varphi_1$) | -0.0140** (0.0066) | -0.4072*** (0.1025) | CDS | -0.0855*** (0.0195) | - |
| $EX_{t-1}$ ($\varphi_2$) | -1.2780*** (0.3312) | 117.2774*** (42.6652) | EX | -7.7974*** (1.8202) | 288.0093*** (62.4297) |
| $CPI_{t-1}$ ($\varphi_3$) | -0.2261* (0.1356) | 5.4331 (3.6444) | CPI | -1.3817 (0.9428) | 13.3426** (6.2251) |
| $INT_{t-1}$ ($\varphi_4$) | -0.0584* (0.0302) | 3.2827* (1.9357) | INT | -0.3586* (0.1898) | 8.0616 (7.3502) |
| $DPOL_{t-1}$ ($\varphi_5$) | -0.0882*** (0.0251) | 36.4096*** (11.1957) | DPOL | -0.5380*** (0.1427) | 89.4153*** (15.4522) |
| $BIST_{t-1}$ ($\varphi_6$) | -0.1639*** (0.0641) | -1.8820*** (0.5203) | BIST | - | -44.0604*** (11.2710) |
| R-Square | 0.4539 | 0.5059 | | | |
| Walt test $\varphi_i=0$ | 4.3927*** | 4.2055** | | | |
| Critical Value Case I | Lower Bound | Upper Bound | | | |
| 10% significance | 1.81 | 2.93 | | | |
| 5% significance | 2.14 | 3.34 | | | |
| 1% significance | 2.82 | 4.21 | | | |

**Notes:** The numbers in the table are coefficients estimated by ARDL analisys technique where significance levels follows as *:10%, **:5%, and ***:1%. Critical values for bound test is retrived from Pesaran et al (2001) Table CI at Case I with k=5.

Further, we investigate short-run dynamics of BIST and CDS using restricted error-correction model defined at equation 2.2a and 2.2b respectively. This technique also provides evidence for how quickly cointegrated series converge to their long-run equilibrium.

The table 9 demonstrates results obtained from this analysis, where coefficients of independent variables imply their short-run causality on dependent one, and coefficient of $ECT_{t-1}$ indicates the speed of error correction. The table shows that both ECTs are negative (between -1 and 0) and statistically significant. This implies that both models do not have instability problems caused by structural break in data. The $ECT_{t-1}$ implies that (2.2a) model corrects 20.18% of its previous month disequilibrium in current month. Meantime, the diagnostic tests confirm that the model is flawless. The estimated results also reveal that 1 TL increase in CDS prices causes 22.5 TL reduction in BIST-100 index in short-run at 1% significance level. Likewise, a kurush increase in TL/USD exchange rate shrinks BIST-100 index by 24.5 TL in short-run. Apparently, current periods CPI seems to be statistically insignificant, however, its first lag (previous period CPI) is statistically significant at 5% level. A percentage increase in lagged inflation rate and current interest rate diminishes BIST-100 index price by 166 TL and 258 TL respectively in short-run. Plus, each negative political event causes BIST-100 index by 237 TL in short-run at 1% significance level.

On the other hand the model 2.2b also predicts plausible results. It estimates that 1000 TL increase in current and lagged BIST-100 index causes jointly 38 TL reducation in current CDS prices in short-run at 1% significance level. Besides, a kurush increase in current TL/USD exchange rate motivates CDS prices by 2.5 TL in short-run, while a percentage increase in inflation rate increases CDS prices by 6 TL at 10% significance level. The interest rates appear to be statistically insignificant factor in explaining short-run dynamics of CDS prices. In addition, *DPOL* derives positive coefficient of 33.36 at 1% significance level, indicating 33 TL increment in response to each negative political events in short-run. This model has adjustment speed of 37.31%, indicating that it corrects 37.31% of its previous month disequilibrium on current month.

**Table 8.** Results of Bound Test

|  | ΔBIST$_t$ | ΔCDS$_t$ |
| --- | --- | --- |
| ΔBIST$_t$ | - | -24.6944*** (5.5101) |
| ΔBIST$_{t-1}$ | -0.3221*** (0.096) | -13.4232*** (3.4080) |
| ΔCDS$_t$ | -0.0225*** (0.0059) | - |
| ΔEX$_t$ | -2.4591*** (0.6788) | 248.43** (135.3055) |
| ΔCPI$_t$ | 0.2577 (0.3953) | 5.7362* (3.3420) |
| ΔCPI$_{t-1}$ | -0.1661** (0.0779) | - |
| ΔINT$_t$ | -0.2309* (0.1335) | 2.7155 (1.8906) |
| ΔDPOL$_t$ | -0.2374*** (0.0702) | 33.3645*** (11.9873) |
| ECT$_{t-1}$ | -0.2018*** (0.0530) | -0.3731*** (0.0915) |

**Notes:** The numbers in the table are coefficients estimated by restrictive error correction ARDL technique where significance levels follows as *:10%, **:5%, and ***:1%. ECT is lagged residual of long-run models of written in equation 2a and 2b. It shows speed of adjustment of previous period disequilbrium on current period.

## 5. Conclusion

The paper studies both short and long-run interaction between BIST-100 index and CDS prices with daily and monthly periods over January 2008 to May 2015. In daily analysis, both BIST and CDS series are stationarilized by taking their first differences (*ΔBIST* and *ΔCDS* respectively), and subsequently we employ Granger causality test. The results reveal bi-directional causality at 1% significance level, where a percentage increase in today's *ΔCDS* shrinks tomorrow's *ΔBIST* by -0.24%. The inverse impact (running from ΔBIST to ΔCDS) is 1/3 times, indicating a percentage increase in today's *ΔBIST* leads -0.08% reduction in *ΔCDS* in the following day.

In case of monthly analysis, preliminary regression output indicates that 51.27% variations in *ΔBIST* is accounted by CDS prices and four control variables such as TL/USD exchange rate, inflation rate, interest rate, and *DPOL* (dummy for negative political events). Estimates show that CDS, exchange rate, interest rate, and *DPOL* appear statistically significant factor in explaining BIST index price. The estimates indicate that 1 TL increase in CDS shrinks BIST index price by 22 TL, while a kurush increase in exchange rate of TL/USD causes 50.4 TL reduction in BIST index price. Likewise, we find that a percentage increase in interest rate decreases BIST index price by 338 TL at 10% significance level. Plus, each negative political event reduces BIST index price by 242 TL at 1% significance level.

In *ΔCDS*-dependent model, BIST and the four control variables accounts 58.53% variations in CDS. This model predicts significant negative impact running from *ΔBIST* to *ΔCDS* (-35.73 in *ΔINT* case and -36.93 in *ΔCPI* case) at 1% significance level. This indicates that 1000 TL increase in BIST index reduces *ΔCDS* price by 37 TL. Besides, a kurush (1/100 TL) increase in exchange rate scales *ΔCDS* up by 2.93 TL (in *ΔINT* case) or 3.09 TL (in *ΔCPI* case). Interestingly, inflation becomes a significant explanatory variable of *ΔCDS* that contribute by additional 5% of $R^2$ value, while interest rate fails to be statistically significant. 1 bps increase in *Δ*CPI causes *Δ*CDS increase by 4.57 TL. Plus, the coefficient of *DPOL* is predicted as -36 at 1% significance level. It indicates that each political uncertainty scales the CDS prices up by 36 TL. These findings reveal that both BIST-100 index and Turkey's CDS prices are severe responsive to the changes in exchange rates and political instability.

Besides, the paper proceeds an ARDL approach to examine both shrot-run and long-run interaction of these series considering, on one hand, *ΔBIST*-dependent (2.1a) model where 45.39% of its variation is captured, and on the other hand, *ΔCDS*-dependent (2.1b) model with 50.59% explanatory power. More specifically, the results of ARDL analysis generate several findings. First, it finds a significant and inverse long-run causality running from *ΔCDS* to *ΔBIST* with impact magnitude of -0.0855 which indicates 1 TL increase in CDS prices pulls down BIST-100 index price by 85.5 TL in long-run at 1% significance level. The restrictive error-correction model shows that this impact is limited with 22.5 TL in short-run.

The system also predicts inverse impact that runs from BIST to CDS as -44.06 in long-run and as -24.69 in short-run. This indicates that 1000 TL increase in BIST index price causes 25 TL and 44 TL reducation in Turkey's CDS prices in short- and long-run respectively. Interestingly, interest rate appears to be significant factor in explaining BIST-100 index, while inflation rate is significant factor for CDS prices, but not vice-versa. This indicate that a percentage increase in interest rate shrinks BIST index by 359 TL and a percentage increase in inflation rate scales CDS prices up to 13.34 TL both in long-run. In case of short-run, these impacts are limited with 231 TL and 5.74 TL respectively.

More importantly, estimates of *DPOL* indicate that each negative political event causes 538 TL reduction in BIST-100 index price and 89 TL increase in CDS prices in long-run at 1% significance level. Restrictive error-correction model implies that these impacts are limited with 237 TL and 33 TL respectively at 1% significance level. This clearly provides strong evidence for severe sensitivity of financial markets to uncertainties in political environment. Plus, both error-correction systems seem to work properly. The *ΔBIST*-dependent (2.2a) model corrects 20.18% of its previous month disequilibrium on current month at 1% significance level, while this adjustment speed is 37.31% in *ΔCDS*-dependent (2.2b) model. The results indicate that the series in (2.2a) and (2.2b) models cannot drift too far apart, and converge to a long-run equilibrium at a moderate monthly speed.

**APPENDIX-A**

| Month-Date | Events with Huge Negative Politic Impact |
|---|---|
| January 2008 | Ergenekon Operations: Initial Waves |
| February 2008 | Ergenekon Operations: Initial Waves continues |
| March 2008 | Ergenekon Operations: Initial Waves continues |
| July 2008 | Closure Trial of AKP (running government) |
| January 2009 | Ergenekon Operations: Latest Waves |
| February 2009 | Ergenekon Operations: Latest Waves continues |
| March 2009 | Local Elections |
| June 2010 | Gaza Aid Flotilla (Mavi Marmara) raid |
| September 2010 | Constitutional Referendum |
| June 2011 | General Elections |
| August 2011 | The Oath Crisis |
| December 2011 | The Roboski (Uludere) airstrike |
| February 2012 | MIT Crisis |
| May 2013 | The Gezi Uprising |
| June 2013 | The Gezi Uprising |
| July 2013 | The Gezi Uprising |
| August 2013 | The Gezi Uprising |
| November 2013 | The AKP-Gulen split |
| December 2013 | The 17-25 December Corruption Operations |
| January 2014 | The Massive Polis Reshuffling & Cassette Scandals |
| March 2014 | Local Elections |
| August 2014 | Presidential Elections |
| September 2014 | Assault on Kobane (Kurdish Protests) |
| December 2014 | Assault on Kobane (Kurdish Protests) |
| February 2015 | The Seizure of Bankasya |
| March 2015 | The Prosecutor Hostage Crisis (Death of Mehmet Selim Kiraz) |